\documentclass[11pt]{article}
\usepackage{amsmath}

\usepackage{graphicx,amssymb,lineno}

\def\ve{\varepsilon}
\def\t#1{\tilde #1}

\def\p{\partial}                                
\newcommand{\Aslash}{A\hspace{-2.0mm}/}         
\newcommand{\pslash}{\partial\hspace{-2.0mm}/}  
\def\slashchar#1{\setbox0=\hbox{$#1$}           
   \dimen0=\wd0                                     
   \setbox1=\hbox{/} \dimen1=\wd1                   
   \ifdim\dimen0>\dimen1                            
      \rlap{\hbox to \dimen0{\hfil/\hfil}}          
      #1                                            
   \else                                            
      \rlap{\hbox to \dimen1{\hfil$#1$\hfil}}       
      /                                             
   \fi}

\def\hc{{\ + \ {\rm h.c.}} }

\newcommand{\initiate}{\setcounter{equation}{0}}

\newcommand{\beq}{\begin{equation}}
\newcommand{\eeq}{\end{equation}}
\newcommand\be{\begin{equation} }
\newcommand\bea{\begin{eqnarray}}
\newcommand\ee{\end{equation}}
\newcommand\eea{\end{eqnarray}}

\def\endtitle{\par\end{quotation}\vskip3.5in minus2.3in\newpage}

\def\a{\alpha}       \def\b{\beta}

       \def\l{\lambda}
\def\m{\mu}          \def\n{\nu}
       \def\r{\rho}
\def\s{\sigma}       \def\t{\tau}

\def\cL{{\mathcal{L} }}

\usepackage{a4wide}
\begin{document}

\title{Chiral fermions in noncommutative electrodynamics:
renormalizability and dispersion}
\vskip75pt
\author{Maja Buri\'c$^{1}$\thanks{majab@ipb.ac.rs},
Du\v sko Latas$^{1}$\thanks{latas@ipb.ac.rs}, Voja Radovanovi\' c$^{1}$\thanks{rvoja@ipb.ac.rs}\\
and Josip Trampeti\'c$^{2}$\thanks{josipt@rex.irb.hr}
                   \\[35pt]$\strut^{1}$
University of Belgrade, Faculty of Physics, \\ P.O.Box 368, RS-11001
Belgrade, Serbia
                   \\[10pt]$\strut^{2}$
      Rudjer Bo\v skovi\' c Institute, Physics Division, \\
P.O.Box 180, HR-10002 Zagreb, Croatia
     }
\date{}
\maketitle
\parindent 0pt
\vskip25pt


\begin{abstract}
We analyze quantization of noncommutative chiral electrodynamics
in the enveloping algebra formalism
in linear order in noncommutativity parameter $\theta$. Calculations show
that divergences exist and cannot be removed by ordinary renormalization,
however they can be removed by the Seiberg-Witten redefinition of fields.
Performing redefinitions explicitly,
we obtain renormalizable lagrangian and discuss the influence of
noncommutativity on field propagation.
Noncommutativity affects the propagation of chiral fermions only:
half of the fermionic modes become massive and birefringent.
\end{abstract}

\maketitle

\section{Introduction}

The original motivation to introduce noncommutativity in the forties~\cite{Snyder:1946qz}
was regularization of divergences in quantum field theory; elimination of singularities
in classical field theories, in particular in gravity adjoined  as a motive
shortly. Till the present day however the program of renormalization through
noncommutativity has not been fully carried out. Initial enthusiasm, when
the subject was reopened in the nineties, decreased after negative results
on renormalizability in the models defined by replacement of the ordinary product
by the Moyal product~\cite{Chepelev:1999tt,Matusis:2000jf,Hayakawa:1999yt}. Research
afterwards diversed in many directions: various modifications of field actions,
different representations of symmetry, new background manifolds were analyzed.
The present status is that we understand properties
of gauge and scalar fields in many details while there is still no general agreement
on how noncommutative gravity should be described.  There are in addition
affirmative results on renormalizability of some particular models.

In their usual versions noncommutative field theories violate Lorentz invariance,
 the corresponding effects are related to the magnitude of noncommutativity
$\theta$. Since there are currently  many experimental searches for relativity 
violations, both in laboratory experiments and in astrophysical measurements,
a straightforward task is to use the known data
to estimate the value of noncommutativity parameters and to probe various
models. As indeed, a priori it is not clear that
 noncommutative field theories can provide with correct models which
would explain some of the observed phenomena.

A possible mechanism to describe the Lorenz violation
is a modification of dispersion relations at high energies.
Of particular interest is the dispersion of photons, as there is a
lot of data which can be used to test it. Modifications due to
noncommutativity were discussed in the literature before
\cite{Carroll:2001ws,Brandt:2002if,Mariz:2006kp,Zahn:2006mg,Abel:2006wj,Bietenholz:2008tp},
both at the classical level and accounting the leading quantum corrections.
The novel feature in this paper is that we discuss it within a renormalizable model;
moreover, the modification is a consequence of the requirement of renormalizability.
Namely, quantizing noncommutative chiral electrodynamics we obtain
that all $n$-point functions, including the propagators, get divergent
contributions. But, compared to the usual procedure our
model allows an additional possibility to yield
renormalizability: the Seiberg-Witten (SW) redefinition of 
fields~\cite{Seiberg:1999vs},  which in principle changes
the form of the lagrangian including the kinetic terms.
In fact renormalizability of other well established models like
the Grosse-Wulkenhaar model~\cite{Grosse:2004yu} was achieved in a
similar manner, by changing the propagator. The result which we  obtain
is that the additional kinetic term in the gauge field action
has  no effect and the photons propagate as usual.
But chiral fermions are sensitive to
noncommutativity: half of the spinor modes acquire mass which is
of order  $1/\sqrt{\theta}$ and depends on the direction:
there is vacuum birefringence. This behavior is new and very interesting.

Along with propagators, interaction vertices change, too.
New processes induced by noncommutative interactions can also be used
to estimate values of the parameters in the theory.
We will not discuss them in this paper, leaving this issue for the future work.

The framework which we use is  the `enveloping algebra formalism' or the
`$\theta$-expanded gauge theory'. $\theta$ stands for the value of
the position commutator $\theta^{\mu\nu} $,
\begin{equation}
[x^\mu \stackrel{\star}{,} x^\nu] \equiv x^\mu \star x^\nu - x^\nu
\star x^\mu = \mathrm{i} \theta^{\mu\nu} ,
\label{can}
\end{equation}
this is the relation which defines the flat noncommutative space.
Commutator in (\ref{can}) is the $\star$-commutator given in terms of
the Moyal product of functions,
\begin{equation}
\phi(x)\star \chi(x) = \mathrm{e}^{\frac{\mathrm{i}}{2}
\theta^{\mu\nu}\frac{\partial}{\partial
x^\mu}\frac{\partial}{\partial y^\nu}}\phi(x)\chi(y)|_{y\to x} .
\label{moyal}
\end{equation}
The $\theta^{\mu\nu}$ is a constant dimensionful tensor.
It gives the length scale on which the quantum structure of spacetime becomes
important; its value can be of order of the square of Planck length or larger.
Having in mind smallness of $\theta$, it makes sense to search for
the effects in the leading, linear order in $\theta$.
This was in part a motivation to introduce $\theta$-expansion
in the noncommutative gauge theories originally~\cite{Madore:2000en}.
One should keep in mind however that the issue of convergence of this expansion
and relations to other theories are still open.

 $\theta$-expanded theories  have interesting properties. They are based
on the enlargement of the initial gauge algebra to its enveloping algebra.
This permits possibility to introduce direct products of gauge groups
and different charges for different particles. Further,
it is known that photon self-energy is renormalizable to all orderds in $\theta$ 
using the SW freedom in quadratic and higher orders,~\cite{Bichl:2001cq}. 
Moreover, pure $SU(N)$ gauge theories are perturbatively renormalizable
in $\theta$-linear order without SW redefinition,~\cite{Buric:2005xe,Latas:2007eu}.
Similar holds for the gauge sector of a suitably defined
generalization of the Standard Model,~\cite{Buric:2006wm}.
Still for some time it was believed  that fermions cannot
be successfully incorporated into a renormalizable theory
because of the so-called 4-$\psi$ divergence,~\cite{Wulkenhaar:2001sq}.
Our main motivation to investigate chiral electrodynamics in more details is
the result that 4-$\psi$ divergence is only related to theories with
Dirac fermions, while it vanishes for
 U(1) and SU(2)  theory with  chiral fermions,~\cite{Buric:2007ix}.
This result was generalized to arbitrary GUT-inspired models with
chiral fermions in~\cite{Martin:2009sg}, and it opened a  possibility 
to construct renormalizable gauge theories with matter. First  such
models were proposed in~\cite{Martin:2009vg,Tamarit:2009iy}, 
where their on-shell one-loop renormalizability was shown. 

In this paper we continue along the same line of investigation by analyzing off-shell 
one-loop renormalizability of noncommutative chiral electrodynamics
in linear order in $\theta$. The results of the calculation show
that divergences exist and that they cannot be removed by ordinary renormalization 
of coupling constants. However, divergences are of the type which can be removed 
by the Seiberg-Witten redefinition of  fields. We perform this redefinition explicitly
and analyze the modification of the propagation properties of fields in our model.

\initiate
\section{Noncommutative chiral electrodynamics}

The commutative action for chiral electrodynamics is
given by
\begin{equation}
S_{\rm C}
= \int \mathrm{d}^4x \left( \mathrm{i}\bar\varphi
\bar\sigma^\mu (\partial_\mu +\mathrm{i}qA_\mu )\varphi -
\frac{1}{4}F_{\mu\nu}F^{\mu\nu} \right),
\label{Scom}
\end{equation}
where  $\varphi$ denotes the left chiral fermion, $q$ is its charge,
 $A_\mu$ is the $\mathrm{U}(1)$ vector potential and
 $F_{\mu\nu} =\p_\mu A_\nu - \p_\nu A_\mu$ is the corresponding field strength.
The noncommutative $\mathrm{U}(1)$ symmetry can be realized by an
analogous set of fields which we denote by a hat:
$\,\hat \varphi\,$,  $\,\hat A_\mu\, $ and  $\,\hat F_{\mu\nu} \,$. As
 noncommutative $\mathrm{U}(1)$ group is  nonabelian,  the field
strength is $\hat F_{\mu\nu} = \partial_\mu \hat A_\nu - \partial_\nu \hat
A_\mu +\mathrm{i}q[\hat A_\mu \stackrel{\star}{,} \hat A_\nu]$;
otherwise all definitions in the two cases are analogous.

Commutative and noncommutative symmetries which correspond to the same gauge group
can be related by the Seiberg-Witten map: the map
gives an explicit relation between corresponding gauge and matter fields.
SW map can also be seen as an expansion in $\theta^{\mu\nu}$
\be
\hat
A_\m=\sum A_\m^{(n)} ,\quad \hat \varphi =\sum \varphi^{(n)} ,
\label{sw}
\ee
where terms $ A_\m^{(n)}$ and $\varphi^{(n)} $ contain
$\theta^{\mu\nu}$ to the $n$-th power. One also assumes that
$\, A_\m^{(0)}= A_\m$,
$\,\varphi^{(0)}= \varphi $, which
fixes the commutative limit $\,\theta^{\mu\nu} =0\,$ of the theory.

Seiberg-Witten expansion (\ref{sw}) can be seen as a solution to
the group closure equations. The simplest
 solution to  linear order is~\cite{Madore:2000en,Schupp:2002up}:
\begin{eqnarray}
&&
\hat A_\rho = A_\rho  +\frac 14 q\, \theta^{\mu\nu} \{ A_\mu,
\partial_\nu A_\rho   + F_{\nu \rho} \} ,
\label{expansion:A}
\\ &&
\hat F_{\rho\sigma} = F_{\rho\sigma}  -\frac{1}{2}q\, \theta^{\mu\nu}
\{ F_{\mu\rho} ,F_{\nu\sigma} \}   + \frac{1}{4} q\, \theta^{\mu\nu}
\{ A _\mu, (\partial_\nu + {D}_\nu
)F_{\rho\sigma}  \} ,
\label{expansion:F}\\
&&
\hat\varphi = \varphi
+\frac{1}{2}q\, \theta^{\mu\nu} A_\mu \partial_\nu\varphi ,
\label{expansion:psi}
\end{eqnarray}
where ${D}_\mu$ denotes the commutative covariant derivative,
$\, D_\m \varphi= (\partial_\mu +\mathrm{i}qA_\mu)\varphi  \, $.
However, this solution is not unique.
It was shown in~\cite{Asakawa:1999cu,Bichl:2001cq}  that  a whole
class of solutions can be obtained from
(\ref{expansion:A}-\ref{expansion:psi}) by a shift of fields
\be\label{red}
{A_\m^{(n)}}\to A_\m^{(n)} + {\bf A}_\m^{(n)},\quad
{\varphi^{(n)}}\to \varphi ^{(n)} + {\bf \Phi}^{(n)} ,
\ee
where $ {\bf A}_\m^{(n)}$ and ${\bf \Phi}^{(n)} $ are
arbitrary gauge covariant expressions
of given order $n$, $\, n>0$.
This means that, if we  assume that noncommutative fields
$\,\hat A_\m\,$ and $\,\hat\varphi\,$ are  primary or `physical'
objects  in the theory and likewise, that noncommutative action is
fixed by a first principle, when
written in commutative fields $ \,A_\m$, $\,\varphi$
the action is not unique. One needs an
additional criterion to decide which of the induced commutative
actions is physical. At the same time, nonuniqueness  gives a new
family of counterterms which can be
used to achieve renormalizability of the theory.

The action for noncommutative chiral electrodynamics is given by
\begin{equation}
S_{\rm NC} =
\int \mathrm{d}^4x \left( \mathrm{i}\hat{\bar \varphi}
\star \bar\sigma^\mu (\partial_\mu +\mathrm{i}q\hat A_\mu ) \star
\hat\varphi- \frac{1}{4}\hat F_{\mu\nu}\star \hat
F^{\mu\nu}\right).
\label{act}
\end{equation}
This action can be expanded in commutative fields
and then quantized by the usual methods. We  truncate the expansion
at linear order in $\theta$.
Using (\ref{expansion:A}-\ref{expansion:psi}) we obtain
\begin{equation}
\mathcal{L}_{\rm NC} = \mathcal{L}_0 +\mathcal{L}_{1,A}
+\mathcal{L}_{1,\varphi}~, \label{lag}
\end{equation}
with
\begin{eqnarray}
&& \mathcal{L}_0 \equiv \mathcal{L}_{\rm C} =
\mathrm{i}\bar\varphi\bar\sigma^\mu( D_\mu
\varphi ) -
\frac{1}{4}F_{\mu\nu}F^{\mu\nu},
\label{L0} \\[8pt]
&&
\mathcal{L}_{1,A} = \frac{1}{2}q\,\theta^{\mu\nu}
\Big( F_{\mu\rho}F_{\nu\sigma}F^{\rho\sigma} -\frac{1}{4}
F_{\mu\nu}F_{\rho\sigma}F^{\rho\sigma}\Big),
\label{L1A} \\[8pt]
&&
\mathcal{L}_{1,\varphi} =
\frac{\rm i}{4}q \Big( \theta^{\mu\nu} F_{\mu\nu}\,\bar\varphi {\bar\sigma}^\r (D_\r \varphi)
-2 \theta^{\mu\nu}F_{\mu\rho}\,\bar\varphi {\bar\s}^\rho(D_\nu \varphi)
\Big)                         \label{**}   \\[8pt]
&&\phantom{\mathcal{L}_{1,\varphi}} =
 \frac{\rm i}{16}q\, \theta^{\mu\nu}
\Delta^{\alpha\beta\gamma}_{\mu\nu\rho}\, F_{\alpha\beta} \,\bar\varphi\,
\bar\sigma^\rho ( D_\gamma
\varphi   ) \hc  \,.
\label{L1phi}
\end{eqnarray}
Cyclic and antisymmetric $\Delta$ in (\ref{L1phi}) is defined by
$\Delta^{\alpha\beta\gamma}_{\mu\nu\rho}=
-\varepsilon^{\alpha\beta\gamma\delta}\varepsilon_{\mu\nu\rho\delta}
$. Though (\ref{lag}) is not unique we start with it because
it is the simplest of the actions. In principle it is possible to treat
the whole class of actions from the beginning (e.g. it was done for a
similar model in~\cite{Martin:2006gw}), but
such approach introduces a large number of coupling constants
and makes an already difficult calculation very complicated.

\initiate
\section{Quantization}

We quantize action~(\ref{act}) by using the path integral
method. Concrete details of the method and of our
notation can be found in~\cite{Buric:2002gm,Buric:2004ms,Buric:2007ix},
 we will stress here
only some specific points. In principle, $\theta$-dependent terms are treated
as interactions and $\theta^{\m\n}$ as a coupling constant. Since the interaction terms
in (\ref{lag}) contain three and more fields,
propagators for the spinor and for the gauge fields are the same
as in commutative theory. To compute the functional
integral one has to complexify the gauge potential or to
introduce the Majorana spinors instead of the chiral; we do the latter.
Denoting the Majorana spinor by $\psi$,
\begin{equation}
 \psi =
\left(
    \begin{array}{c}
        \varphi_\alpha \\
       \bar\varphi^{{\dot\alpha}}
    \end{array}
\right) ,
\end{equation}
we can write the commutative part of the lagrangian
in the form
\begin{equation}
\mathcal{L}_0  =
\frac{\mathrm{i}}{2}\,\bar\psi\gamma^\mu(\partial_\mu
-\mathrm{i}q\gamma_5 A_\mu )\psi -
\frac{1}{4}F_{\mu\nu}F^{\mu\nu}. \label{L0'}
\end{equation}
Expressions
(\ref{L0'}) and (\ref{L0}) are identical in the chiral representation
of $\gamma$-matrices (\ref{gama}).  The
$\theta$-linear spinor part of the lagrangian is expressed as
\begin{equation}
\mathcal{L}_{1,\varphi} =\frac{\rm i}{16}q \,\theta^{\mu\nu}
\Delta^{\alpha\beta\gamma}_{\mu\nu\rho} \, F_{\alpha\beta}
\bar\psi\gamma^\rho(\partial_\gamma
-\mathrm{i}q\gamma_5A_\gamma)\psi,
\label{L1psi}
\end{equation}
while the gauge part is of course the same, (\ref{L1A}).

In order to preserve the gauge covariance we  use the
background field method. Briefly: we
expand fields around their classical configurations, replacing formally
 in the action $\,A_\mu\to  A_\mu + \mathcal{A}_\mu\,$ and $ \,\psi \to \psi + \Psi
\,$; then we integrate over the quantum fields
 $\,\mathcal{A}_\mu\,$, $\Psi\,$. After the integration we obtain the one-loop
effective action $\Gamma$,
\begin{equation}
\Gamma[A_\mu,\psi] = S_{cl}[A_\mu,\psi]
-\frac{1}{2\mathrm{i}}\,\mathrm{STr} \log \mathcal{B}
[A_\mu,\psi] .
\label{*}
\end{equation}
The first term is the classical part,  the
second term is the one-loop quantum correction. Operator
$\mathcal{B} $, the result of Gaussian integration,
 can be obtained as the term of second order in the expansion of classical action
$\, S_{cl}[A_\mu+\mathcal{A}_\mu,\psi+\Psi]\, $ in fields $\mathcal{A}_\mu\,$ and
$\,\Psi$,
\begin{equation}
S^{(2)} = \int \mathrm{d}^4x
\begin{pmatrix}
    \mathcal{A}_\kappa & \bar\Psi
\end{pmatrix}\,
\mathcal{B}\,
\begin{pmatrix}
    \mathcal{A}_\lambda \\[4pt]
    \Psi
\end{pmatrix}.
\end{equation}

$\mathcal{B}$ can be divided into commutative part
$\mathcal{B}_0$ and  $\theta$-linear part
$\mathcal{B}_1$, $\,\mathcal{B} = \mathcal{B}_0 +\mathcal{B}_1\,$.
 After inclusion of the gauge fixing terms, $\mathcal{B}_0$  is given by
\begin{equation}
\mathcal{B}_0 = \frac{1}{2}
\begin{pmatrix}
    g^{\kappa\lambda}\Box & q\bar\psi \gamma^\kappa \gamma_5 \\[4pt]
    q\gamma^\lambda \gamma_5 \psi & \mathrm{i}\pslash +q\Aslash \gamma_5
\end{pmatrix} ;
\label{B0}
\end{equation}
it has the kinetic part
$\mathcal{B}_\mathrm{kin} =${\scriptsize$
 \frac{1}{2}
\begin{pmatrix}
    g^{\kappa\lambda}\Box &0 \\[4pt]
   0 & \mathrm{i}\pslash
\end{pmatrix}
$}
and the interaction.

In order to calculate the one-loop effective action perturbatively
we need to expand the logarithm in~(\ref{*})
around identity
$\mathcal{I} =${\scriptsize$
\begin{pmatrix}
    g^{\kappa\lambda} & 0 \\[4pt]
    0 & 1
\end{pmatrix} $}.
Thus  we have to multiply $\mathcal{B}$ by  matrix
$\mathcal{C}$,
\begin{equation}
\mathcal{C} = 2
\begin{pmatrix}
    g^{\kappa\lambda} & 0 \\[4pt]
    0 & -\mathrm{i}\pslash
\end{pmatrix} ;
\end{equation}
we then have $ \mathcal{B}_\mathrm{kin}\,\mathcal{C} =\mathcal{I}$ and
\begin{equation}
\mathrm{STr} \left( \log \mathcal{B} \right)= \mathrm{STr}
\left(\log \Box^{-1}\mathcal{BC} \right) - \mathrm{STr}
\left(\mathcal{C} \Box^{-1}\right).              \label{str}
\end{equation}
The first term in (\ref{str}) we denote by $\Gamma^{(1)}$ as,
clearly, up to a constant infinite normalization
$\, \mathrm{STr}
\left(\mathcal{C} \Box^{-1}\right)\,$,  it can be identified
with first quantum correction of the one-loop effective action.

Introducing
\begin{equation}
\mathcal{BC} = \Box \mathcal{I} +N_1 +T_1 +T_2 ,
\label{3parts}
\end{equation}
we obtain the perturbation expansion
\begin{eqnarray}
\Gamma^{(1)} & =& \frac{\mathrm{i}}{2}\,
\mathrm{STr} \log \left(\mathcal{I}
+\Box^{-1}N_1 +\Box^{-1}T_1+\Box^{-1}T_2\right)\nonumber \\[4pt]
&=&\frac{\mathrm{i}}{2} \sum\frac{(-1)^{n+1}}{n}
\mathrm{STr}\left(\Box^{-1}N_1+\Box^{-1}T_1
+\Box^{-1}T_2 \right)^n        .
\label{perturb}
\end{eqnarray}
Interaction  in (\ref{3parts}) is divided in three
parts in the following way. Operator $N_1$ contains
commutative vertices; in case of electrodynamics there is only a 3-vertex,
so $N_1$ contains terms with one classical (external)
and two quantum fields. By analogy, $T_1$ is a term linear in
$\theta$ which contains one classical field and two quantum fields.
$T_2$ is linear in $\theta$ and contains two classical and two quantum
fields. From~(\ref{B0}) we obtain
\begin{eqnarray}
 && N_1 =
q\left(\begin{array}{cc}
   0 & -\mathrm{i}\bar\psi\gamma_5\gamma^\lambda \slashchar{\partial} \\[4pt]
       -\gamma_5\gamma^\kappa\psi & \mathrm{i}\gamma_5\slashchar{A} \slashchar{\partial} \\
\end{array}
    \right),  \\[16pt]
&&T_1 =
-q\left(
      \begin{array}{cc}
      V^{\kappa\lambda} & -\frac{1}{4}\theta^{\mu\nu}\Delta^{\alpha\beta\gamma}_{\mu\nu\rho}
      \delta^\kappa_\alpha (\partial_\beta \bar\psi)\gamma^\rho\partial_\gamma\slashchar{\partial} \\[4pt]
      -\frac{\mathrm{i}}{4}\theta^{\mu\nu}\Delta^{\alpha\beta\gamma}_{\mu\nu\rho}\delta^\lambda_\alpha
   \gamma^\rho(\partial_\beta \psi)\partial_\gamma & -\frac{1}{8}\theta^{\mu\nu}
 \Delta^{\alpha\beta\gamma}_{\mu\nu\rho} F_{\alpha\beta}\gamma^\rho \partial_\gamma \slashchar{\partial} \\
 \end{array}
  \right),
\end{eqnarray}
where the matrix elements $V^{\kappa\lambda} $ are given as $V^{\kappa\lambda}=
-\partial_\sigma V^{\sigma\kappa,\tau\lambda}\partial_\tau\,$ with,~\cite{Buric:2002gm},
\begin{eqnarray}
V^{\sigma\kappa,\tau\lambda} &=&\frac{1}{2}(g^{\sigma\tau} g^{\kappa\lambda}-g^{\sigma\lambda}
g^{\tau\kappa})\theta^{\alpha\beta}F_{\alpha\beta}\nonumber \\
& & - g^{\kappa\lambda}(\theta^{\xi\sigma}{F_\xi}^\tau+\theta^{\xi\tau}{F_\xi}^\sigma) -
g^{\sigma\tau}(\theta^{\xi\kappa}{F_\xi}^\lambda+\theta^{\xi\lambda}{F_\xi}^\kappa)\nonumber \\[4pt]
& & + g^{\kappa\tau}(\theta^{\xi\lambda}{F_\xi}^\sigma+\theta^{\xi\sigma}{F_\xi}^\lambda) +
g^{\sigma\lambda}(\theta^{\xi\kappa}{F_\xi}^\tau+\theta^{\xi\tau}{F_\xi}^\kappa)\nonumber \\[4pt]
& & -\theta^{\kappa\lambda}F^{\sigma\tau} + \theta^{\kappa\tau}F^{\sigma\lambda} +
\theta^{\sigma\lambda}F^{\kappa\tau} + \theta^{\sigma\kappa}F^{\tau\lambda} +
\theta^{\tau\lambda}F^{\sigma\kappa} - \theta^{\sigma\tau}F^{\kappa\lambda} .
\end{eqnarray}
$T_2$ is equal to
\begin{equation}
 T_2 = \frac{1}{8}q^2\theta^{\mu\nu}\Delta^{\alpha\beta\gamma}_{\mu\nu\rho}
\left(
     \begin{array}{cc}
     \delta^\kappa_\alpha \delta^\lambda_\beta (\partial_\gamma \bar\psi\gamma_5\gamma^\rho\psi
     +\bar\psi \gamma_5\gamma^\rho \psi\partial_\gamma)\  & \mathrm{i} \delta^\kappa_\alpha
(2\partial_\beta A_\gamma + F_{\beta\gamma}) \bar\psi\gamma_5\gamma^\rho \slashchar{\partial} \\[4pt]
  \delta^\lambda_\alpha \gamma_5\gamma^\rho\psi (2A_\gamma \partial_\beta-F_{\beta\gamma}) &
  \mathrm{i} F_{\alpha\beta}A_\gamma\gamma_5\gamma^\rho\slashchar{\partial} \\
  \end{array}
  \right).
\end{equation}
The order of  operators
is of importance;  in our notation we have for example 
\begin{eqnarray}
&&\Delta^{\alpha\beta\gamma}_{\mu\nu\rho}\,\partial_\beta A_\gamma=\Delta^{\alpha\beta\gamma}_{\mu\nu\rho}\,\Big((\partial_\beta A_\gamma)+A_\gamma\partial_\beta\Big)= \,\Delta^{\alpha\beta\gamma}_{\mu\nu\rho}\Big( \frac 12 F_{\beta\gamma}+A_\gamma\partial_\beta\Big),\\[4pt]
&&
\partial_\gamma\bar\psi\gamma_\rho\gamma_5\psi = (\partial_\gamma(\bar\psi\gamma_\rho\gamma_5\psi))
+\bar\psi\gamma_\rho\gamma_5\psi \partial_\gamma
\nonumber
\end{eqnarray}
and so on.

\initiate
\section{Divergences and renormalization}

We would like to extract divergent parts of the one-loop effective action from
expansion (\ref{perturb}); the relevant terms can be identified by power counting.
Divergences exist in the 2-point, 3-point and 4-point functions and
a careful analysis shows that they  are contained  only in  terms
$\mathrm{STr}(\Box^{-1}N_1\Box^{-1}T_1)\,$, 
$\mathrm{STr}(\Box^{-1}N_1\Box^{-1}T_2)\,$
and $\mathrm{STr}(\Box^{-1}N_1\Box^{-1}N_1\Box^{-1}T_1)\,$.
Supertraces can be calculated in a standard fashion, in the momentum
representation using dimensional regularization.
The calculation itself however is very demanding and to obtain
the results we combined  ordinary calculation with the algebraic one using
the {\it Math\-Ten\-sor} package in {\it Mathematica}.
To separate contributions coming  from different $n$-point functions we
denote
\begin{equation}
{\Gamma^{(1)}}|_{\mathrm{div}} =
{{\it\Gamma}_2} + {{\it\Gamma}_3} + {{\it\Gamma}_4} = \Gamma_2 + \Gamma_3 .
\end{equation}
$\Gamma_2$ and $\Gamma_3$ will be used later and they denote divergences written 
in a covariant form.

The divergent part of  the 2-point function is:
\begin{eqnarray}
&&
{\it\Gamma}_2 = -\frac{\rm i}{2}\, \mathrm{STr}(\Box^{-1}N_1\Box^{-1}T_1)|_{\mathrm{div}}\label{Gamma2:it} \\[4pt]
&&\phantom{{\it\Gamma}_2 }= -\frac{\rm i}{8}q^2 \theta^{\mu\nu} \Delta^{\alpha\beta\gamma}_{\mu\nu\rho}\, \mathrm{Tr}\left(\Box^{-1}\bar\psi \gamma_5 \gamma_\alpha \slashchar{\partial}\Box^{-1} \gamma^\rho (\partial_\gamma \psi) \partial_\beta + \Box^{-1}\gamma_5 \gamma_\alpha \psi\Box^{-1} (\partial_\beta \bar\psi) \gamma^\rho \partial_\gamma\slashchar{\partial}\right)\nonumber \\[4pt]
&&\phantom{{\it\Gamma}_2  =}
-\frac{1}{16}q^2 \theta^{\mu\nu} \Delta^{\alpha\beta\gamma}_{\mu\nu\rho}\, \mathrm{Tr}\left( \Box^{-1} \gamma_5 \slashchar{A} \slashchar{\partial} \Box^{-1} F_{\alpha\beta}\gamma^\rho \partial_\gamma\slashchar{\partial}\right)\nonumber \\[4pt]
&&\phantom{{\it\Gamma}_2 }= -\frac{\rm i}{48} \frac{q^2}{(4\pi)^2 \epsilon} \theta^{\mu\nu} \Delta^{\alpha\beta\gamma}_{\mu\nu\rho} (\partial_\gamma \bar\psi) \left( \gamma_5 \gamma_\alpha \gamma_\beta \gamma^\rho -  \gamma_5 \gamma^\rho\gamma_\beta \gamma_\alpha \right) (\Box \psi) \nonumber \\[4pt]
&&\phantom{{\it\Gamma}_2  = } - \frac{\mathrm{i}}{12} \frac{q^2}{(4\pi)^2 \epsilon} \theta^{\mu\nu} \Delta^{\alpha\beta\gamma}_{\mu\nu\rho} \varepsilon^{\sigma\kappa\lambda\rho} A_\kappa \eta_{\lambda\gamma} (\partial_\sigma \Box F_{\alpha\beta}) ,   \nonumber
\end{eqnarray}
so we obtain
\begin{equation}
\hskip-40pt
{\it\Gamma}_2 =
- \frac{1}{12} \frac{q^2}{(4\pi)^2\epsilon}\, \theta^{\mu\nu}
\Big( {\mathrm{i}}
{\varepsilon_{\mu\nu}}^{\rho\sigma}(\partial_\rho\bar\psi)\gamma_\sigma
(\Box \psi) + {{\varepsilon_{\mu}}^{\rho\sigma\tau}}
F_{\rho\sigma}(\Box F_{\nu\tau})  \Big).
\label{G2Maj}
\end{equation}

Calculation of the divergent parts of the 3-point functions gives:
\begin{eqnarray}
&&\left.\mathrm{STr}(\Box^{-1}N_1\Box^{-1}T_2)\right|_{\mathrm{div}} =- \frac{q^3}{(4\pi)^2\epsilon} \theta^{\mu\nu} \left( -\frac{\rm i}{3}A_\rho (\partial_\mu \bar\psi) \gamma^\rho (\partial_\nu \psi) - \frac{\rm i}{3}A_\mu (\partial_\nu \bar\psi) \gamma^\rho (\partial_\rho \psi) \right.  \nonumber \\[4pt]
&& \phantom{\quad\quad}
- \frac{\rm i}{3}A^\rho (\partial_\mu \bar\psi) \gamma_\nu (\partial_\rho \psi) - \frac{1}{3}{\varepsilon_{\mu}}^{\rho\sigma\tau} A_\rho (\partial_\nu \bar\psi) \gamma_5 \gamma_\sigma (\partial_\tau \psi) -\frac{4 \rm i}{3} F_{\mu\rho}\bar\psi \gamma^\rho (\partial_\nu \psi)
\nonumber \\[4pt]
&&\phantom{\quad\quad}
-\frac{4 \rm i}{3} F_{\mu\rho}\bar\psi \gamma_\nu (\partial^\rho \psi)
  -\frac{4}{3} {\varepsilon_{\mu}}^{\rho\sigma\tau} F_{\nu\rho} \bar\psi \gamma_5 \gamma_\sigma (\partial_\tau \psi)
-\frac{2\rm i}{3}A_\mu \bar\psi \gamma_\nu (\Box \psi)
\nonumber \\[4pt]
&& \phantom{\quad\quad}
 + \frac{1}{6} {\varepsilon_{\mu\nu}}^{\rho\sigma} A_\rho \bar\psi \gamma_5 \gamma_\sigma (\Box \psi) + \frac{2}{3} A^\rho F_{\mu\nu} (\partial_\sigma F_{\rho\sigma})\left.-\frac{4}{3}A_\mu F_{\nu\rho} (\partial_\sigma F^{\rho\sigma}) \right),\nonumber\\[4pt]
&&
\left.\mathrm{STr}(\Box^{-1}N_1\Box^{-1}N_1\Box^{-1}T_1)\right|_{\mathrm{div}} = -\frac{q^3}{(4\pi)^2\epsilon} \theta^{\mu\nu} \left( -\frac{\rm i}{3}A_\rho (\partial_\mu \bar\psi) \gamma^\rho (\partial_\nu \psi) - \frac{\rm i}{3}A_\mu (\partial_\nu \bar\psi) \gamma^\rho (\partial_\rho \psi) \right.  \nonumber\\[4pt]
&& \phantom{\quad\quad}
- \frac{5 \rm i}{3}A^\rho (\partial_\mu \bar\psi) \gamma_\nu (\partial_\rho \psi) - \frac{1}{3}{\varepsilon_{\mu}}^{\rho\sigma\tau} A_\rho (\partial_\nu \bar\psi) \gamma_5 \gamma_\sigma (\partial_\tau \psi)
 +\frac{\rm i}{3} F_{\mu\rho}\bar\psi \gamma^\rho (\partial_\nu \psi)
\nonumber \\[4pt]
&& \phantom{\quad\quad}
+\frac{2 \rm i}{3} F_{\mu\rho}\bar\psi \gamma_\nu (\partial^\rho \psi) +\frac{\rm i}{6} F_{\mu\nu}\bar\psi \gamma^\rho (\partial_\rho \psi)
-\frac{2 \rm i}{3} (\partial_\rho A^\rho) \bar\psi \gamma_\mu (\partial_\nu\psi)
\nonumber \\[4pt]
&& \phantom{\quad\quad}
 + \frac{1}{3}{\varepsilon_{\mu\nu}}^{\rho\sigma} A^\tau
(\partial_\rho\bar\psi)\gamma_5 \gamma_\sigma (\partial_\tau\psi)
 -\frac{1}{6}{\varepsilon_{\mu\nu}}^{\rho\sigma} (\partial_\tau
A^\tau) \bar\psi \gamma_5 \gamma_\rho
(\partial_\sigma\psi) +\frac{1}{12}{\varepsilon_{\mu}}^{\rho\sigma\tau} F_{\rho\sigma} \bar\psi \gamma_5 \gamma_\tau
(\partial_\nu\psi)
\nonumber \\[4pt]
&& \phantom{\quad\quad}
 \left. + \frac{1}{6}{\varepsilon_{\mu}}^{\rho\sigma\tau} F_{\nu\rho} \bar\psi \gamma_5 \gamma_\sigma
(\partial_\tau\psi) - \frac{1}{4}{\varepsilon_{\mu\nu}}^{\rho\sigma} F_{\rho\sigma} \bar\psi \gamma_5 \gamma^\tau
(\partial_\tau\psi)  \right) ,\nonumber
\end{eqnarray}
so the result for the divergent 3-vertices is given by:
\begin{eqnarray}
&&\hskip-8pt{\it\Gamma}_3 =
\frac{\rm i}{2}\left(\left.\mathrm{STr}(\Box^{-1}N_1 \Box^{-1} N_1 \Box^{-1} T_1)\right|_{\mathrm{div}}-\left.\mathrm{STr}(\Box^{-1}N_1 \Box^{-1}T_2)\right|_{\mathrm{div}}\right)
\nonumber\\[4pt]
&&\phantom{\Gamma_3} = -\frac{q^3}{(4\pi)^2\epsilon} \theta^{\mu\nu}
\left(\frac{1}{6}F_{\mu\nu}F^{\rho\sigma}F_{\rho\sigma}
-\frac{2}{3}F_{\mu\rho}F^{\nu\sigma}F_{\rho\sigma}+\frac{\mathrm{5i}}{6} F_{\mu\rho}\bar\psi\gamma^\rho
(\partial_\nu\psi) \right.
\label{G3Maj}
\\[4pt]
&&\phantom{\Gamma_3 =}
- \frac{\mathrm{i}}{6}
F_{\mu\rho}\bar\psi\gamma_\nu (\partial^\rho\psi) - \frac{2\mathrm{i}}{3} F_{\mu\nu}\bar\psi\gamma^\rho
(\partial_\rho\psi)+\frac{4}{3}{\varepsilon_{\mu}}^{\rho\sigma\tau}
F_{\rho\sigma} \bar\psi\gamma_5 \gamma_\tau (\partial_\nu\psi)
\nonumber    \\[4pt]
&&\phantom{\Gamma_3 =}
+\frac{3}{2}{\varepsilon_{\mu\nu}}^{\rho\sigma} F_{\rho\tau}
\bar\psi\gamma_5 \gamma_\sigma (\partial^\tau\psi) +\frac{1}{8}{\varepsilon_{\mu\nu}}^{\rho\sigma} F_{\rho\sigma}
\bar\psi\gamma_5 \gamma^\tau
(\partial_\tau\psi)+\frac{1}{12}{\varepsilon_{\mu\nu}}^{\rho\sigma}
A_\rho \bar\psi \gamma_5 \gamma_\sigma
(\Box\psi)
\nonumber \\[4pt]
&&\phantom{\Gamma_3 =}
 \left.-\frac{1}{6}{\varepsilon_{\mu\nu}}^{\rho\sigma} A^\tau
(\partial_\rho\bar\psi)\gamma_5 \gamma_\sigma (\partial_\tau\psi)  +
\frac{1}{12}{\varepsilon_{\mu\nu}}^{\rho\sigma} (\partial_\tau
A^\tau) \bar\psi \gamma_5 \gamma_\rho
(\partial_\sigma\psi)\right) .
\nonumber
\end{eqnarray}

It would be more transparent to have these expressions
written  in covariant derivatives,
however that is not possible in the Majorana representation.
Therefore we rewrite (\ref{G2Maj}) and  (\ref{G3Maj}) in the chiral representation,
collecting together covariant pieces. This mixes
${\it \Gamma}_2$ and ${\it \Gamma}_3$; for example,
the last three terms of (\ref{G3Maj}) belong in fact to the 2-point
function when we write it in covariant derivatives. The divergences become
\begin{equation}
{\Gamma}_2 =\frac{1}{12}\frac{q^2}{(4\pi)^2\epsilon}\theta^{\mu\nu} {\varepsilon_{\mu\rho\sigma\tau}}
(\partial_\lambda F^{\rho\lambda})(\partial_\nu F^{\sigma\tau} )
- \frac{1}{12} \frac{q^2}{(4\pi)^2\epsilon}
 \theta^{\mu\nu}
\varepsilon_{\mu\nu\rho\sigma}\left( \, {\rm i} (D^\rho\bar\varphi){\bar\sigma}^\sigma
(D^2\varphi) \hc \right)   ,
\label{G2}
\end{equation}
and
\begin{eqnarray}
&&\hskip-12pt {\Gamma}_3
= -\frac{q^3}{(4\pi)^2\epsilon} \theta^{\mu\nu}
\left(  \frac{1}{6}F_{\mu\nu}F_{\rho\sigma}F^{\rho\sigma}-
\frac{2}{3}F_{\mu\rho}F_{\nu\sigma}F^{\rho\sigma}  \right)
\label{G3}\\[4pt]
&&\phantom{\Gamma }
- \frac{q^3}{(4\pi)^2\epsilon} \theta^{\mu\nu}
\left( \frac{\mathrm{5i}}{6} F_{\mu\rho}\, \bar\varphi{\bar\sigma}^\rho
(D_\nu\varphi) - \frac{\mathrm{i}}{6}
F_{\mu\rho}\, \bar\varphi{\bar\sigma}_\nu (D^\rho\varphi) -
\frac{2\mathrm{i}}{3} F_{\mu\nu}\, \bar\varphi{\bar\sigma}^\rho
(D_\rho\varphi) \right.
\nonumber\\[4pt]
&&\phantom{\Gamma_3 = \frac{1}{(4\pi)^2\epsilon} \theta^{\mu\nu} }
 +\frac{4}{3} \varepsilon_{\mu\rho\sigma\tau}
F^{\rho\sigma}\, \bar\varphi{\bar\sigma}^\tau (D_\nu\varphi)
+ \frac{3}{2}\varepsilon_{\mu\nu\rho\tau} F^{\rho\sigma}\,
\bar\varphi{\bar\sigma}^\tau (D_\sigma\varphi)
\nonumber \\
&&\phantom{\Gamma_3 + \frac{1}{(4\pi)^2\epsilon}\theta^{\mu\nu} }
 \left.+\frac{1}{8}\varepsilon_{\mu\nu\rho\sigma} F^{\rho\sigma}\,
\bar\varphi{\bar\sigma}^\tau
(D_\tau\varphi) \hc \right)  .
 \nonumber
\end{eqnarray}
The last formula contains all vertex terms, including the 4-point functions ${{\it\Gamma}_4} $.

From the form of (\ref{G2}) and (\ref{G3})
it is quite clear that divergences cannot be removed by multiplicative renormalization:
there are for example divergent contributions to the propagators
while both propagating fields are massless. Also interaction
terms in $\Gamma_3$ have not the form of the interaction lagrangian $\mathcal{L}_{1,\varphi}$.
Therefore the only possibility for renormalization is
the Seiberg-Witten redefinition of fields, and we will explore this possibility more closely.
Of course the analysis which follows is, as all other calculations, done only 
in linear order in $\theta$.

A general SW redefinition (\ref{red}) induces in the action the following additional terms:
\begin{eqnarray}
&& \Delta S^{(n,A)} =\int {\rm d}^4 x (D_\r F^{\r\m}){\bf A}_\m^{(n)} , \label{DSA} \\[4pt]
&& \Delta S^{(n,\varphi)} = {\rm i} \int {\rm d}^4 x \bar\varphi \bar\sigma^\m(D_\m {\bf \Phi}^{(n) })
\label{DSphi} \hc \ ,
\end{eqnarray}
so we need to rewrite $\Gamma_2$ and $\Gamma_3$ in such form,
of course except for the terms proportional to $\mathcal{L}_{1,A}$
and  $\mathcal{L}_{1,\varphi}$.  We can see immediately that the bosonic part of
the two-point divergence is already in the required form. The
shift of the gauge potential
$ A_\rho \to A_\rho +{\bf A}_\rho$, with
\begin{equation}
{\bf A}_\rho = \frac{1}{12} \frac{q^2}{(4\pi)^2\epsilon}\theta^{\mu\nu}
\varepsilon_{\mu\rho\sigma\tau}(D_\nu F^{\sigma\tau})
\label{1}
\end{equation}
cancels it. In fact a similar shift of the spinor $\varphi$,
$\, \varphi\to \varphi +{\bf \Phi} $,  can be done to cancel the fermionic part
of $\Gamma_2$, too.  Using relations (\ref{3sigma})
 among $\s$ and ${\bar\s}$-matrices we obtain that for
\begin{equation}
{\bf \Phi} = -\frac{\rm i}{6} \frac{q^2}{(4\pi)^2\epsilon}\theta^{\mu\nu}
\sigma_{\m\nu}(D^2 \varphi)
\label{2}
\end{equation}
the effective action transforms to
\begin{equation}
 \Gamma = S_{\rm cl} + \Gamma_2 + \Gamma_3 \to S_{\rm cl} + \Gamma_3 + \Gamma_3^\prime ,
\end{equation}
removing $\Gamma_2$ completely.
Redefinitions (\ref{1})-(\ref{2}) along with the cancellation of $\Gamma_2$
induce an additional term $\Gamma_3^\prime$,
\begin{equation}
 \Gamma_3^\prime =-\frac {1}{12}\frac{q^3}{(4\pi)^2\epsilon}\theta^{\m\n}\Big(
2{\rm i} F_{\n\r}{\bar\varphi}{\bar\s}_\m(D_\r \varphi )
-\ve_{\m\r\s\tau}F^{\s\tau} {\bar\varphi}{\bar\s}^\r(D_\n \varphi ) \Big) \hc \,  ,
\end{equation}
and therefore in the next step we have to `redefine away' the 3-point divergence 
\begin{eqnarray}
&&\hskip-12pt {\Gamma}_3+\Gamma_3^\prime
= \frac 43\frac{q^2}{(4\pi)^2\epsilon} \, \cL_{1,A}
\label{G'}\\[4pt]
&&\phantom{\Gamma\qquad }
- \frac{q^3}{(4\pi)^2\epsilon} \theta^{\mu\nu}
\Big( \frac{\mathrm{5i}}{6} F_{\mu\rho}\, \bar\varphi{\bar\sigma}^\rho
(D_\nu\varphi) - \frac{\mathrm{i}}{3}
F_{\mu\rho}\, \bar\varphi{\bar\sigma}_\nu (D^\rho\varphi) -
\frac{2\mathrm{i}}{3} F_{\mu\nu}\, \bar\varphi{\bar\sigma}^\rho
(D_\rho\varphi) \hc \Big)
\nonumber\\[4pt]
&&\phantom{\Gamma\qquad } - \frac{q^3}{(4\pi)^2\epsilon} \theta^{\mu\nu}
F^{\rho\sigma}\, \bar\varphi{\bar\sigma}^\tau
\Big(
\frac{5}{4} \varepsilon_{\mu\rho\sigma\tau}(D_\nu\varphi) + \frac{3}{2} \ve_{\m\n\r\t}(D_\sigma\varphi)
+\frac{1}{8} \ve_{\m\n\r\s}
(D_\tau\varphi)
 \hc \Big)   .
 \nonumber
\end{eqnarray}

At the first sight it looks as if there were too many terms in (\ref{G'}) to cancel: six.
However, by inspecting them separately we can see that this is not the case.
For example it is obvious that the last terms in the second and in the third line,
$ F_{\mu\nu}\, \bar\varphi{\bar\sigma}^\rho
(D_\rho\varphi)$ and $\varepsilon_{\mu\nu\rho\sigma} F^{\rho\sigma}\,
\bar\varphi{\bar\sigma}^\tau
(D_\tau\varphi)  $, are already in the form adjusted for the spinor field
redefinition; this leaves four terms. Let us first discuss the second line of
expression~(\ref{G'}). We observe that it is possible to combine  terms
to obtain $\cL_{1,\varphi}$; for example,
$\, \theta^{\m\n}F_{\m\r}\bar\varphi {\bar\s}^\r(D_\n \varphi)\,$ can be replaced with
\begin{equation}
\theta^{\m\n}F_{\m\r}\bar\varphi {\bar\s}^\r(D_\n \varphi)
= \frac 12 \,\theta^{\m\n} F_{\m\n}\bar\varphi {\bar\s}^\r(D_\r \varphi)
- \frac 14 \,\theta^{\m\n}\Delta^{\a\b\gamma}_{\m\n\r} F_{\a\b}\bar\varphi{\bar\s}^\r(D_\gamma \varphi),
\end{equation}
that is we can use
\begin{equation}
{\rm i} \theta^{\m\n}F_{\m\r}\bar\varphi {\bar\s}^\r(D_\n \varphi)\hc
= \frac {\rm i}{2} \,\theta^{\m\n} F_{\m\n}\bar\varphi {\bar\s}^\r(D_\r \varphi)\hc -
\, \frac{4}{q}\cL_{1,\varphi}  .
        \end{equation}

Therefore in fact there is only one nontrivial term in the second line of (\ref{G'}) and
it can be removed by a spinor shift
\begin{equation}
 {\bf \Phi}^\prime = \frac 23 \frac{q^3}{(4\pi)^2\epsilon}\theta^{\mu\nu}F_{\m\r}\s_{\n\r }\varphi \, .
\label{22}
\end{equation}
This gives additional contribution $\,\Gamma_3^{\prime\prime}\,$ to $\,\Gamma$, 
so after the second shift (\ref{22})  we have
\begin{eqnarray}
&&\hskip-24pt {\Gamma}_3+\Gamma_3^\prime +\Gamma_3^{\prime\prime}
=  \frac 43 \frac{q^2}{(4\pi)^2\epsilon}\,\cL_{1,A}
+ \frac{2q^2}{(4\pi)^2\epsilon}\, \cL_{1,\varphi}
 +\frac{5\rm i}{12}\frac{q^3}{(4\pi)^2\epsilon}\theta^{\mu\nu}
 \Big( F_{\m\n}\bar\varphi{\bar\sigma}^\r(D_\r \varphi) \hc   \Big)
\label{33'3''}\\[4pt]
&&\phantom{\Gamma\quad\quad }
- \frac{q^3}{(4\pi)^2\epsilon} \theta^{\mu\nu}
F^{\rho\sigma}\, \bar\varphi{\bar\sigma}^\tau
\Big(
\frac{17}{12} \varepsilon_{\mu\rho\sigma\tau}(D_\nu\varphi) + \frac 32 \ve_{\m\n\r\t}(D_\sigma\varphi)
+\frac{1}{8} \ve_{\m\n\r\s}
(D_\tau\varphi)
 \hc \Big)   .
 \nonumber
\end{eqnarray}

The last line of formula (\ref{33'3''}) again has  three terms
but only two are independent: this time due to identity (\ref{e-d}). Furthermore,
 $\, \varepsilon_{\mu\nu\rho\tau} F^{\rho\sigma}\,
\bar\varphi{\bar\sigma}^\tau (D_\sigma\varphi) \,$ can be cancelled by a
redefinition of the gauge field ${\bf A}_\rho^\prime $,
\begin{equation}
 {\bf A}_\rho^\prime = -\frac{1}{12}\frac{q^2}{(4\pi)^2\epsilon}
 \, \theta^{\m\n}\ve_{\m\n\r\tau}(\p^\s F_{\tau\s}),
\end{equation}
which  does not change  the bosonic part of the action. Thus after  transormation
$ \, A_\rho\to A_\rho +{\bf A}_\rho +{\bf A}_\rho^{\prime}\,$,
$\, \varphi \to \varphi +{\bf \Phi} +{\bf \Phi}^\prime\, $ we obtain
\begin{eqnarray}
&&\Gamma \to S_{\rm cl} + {\Gamma}_3+\Gamma_3^\prime +\Gamma_3^{\prime\prime} \, =
\label{G'''} \\[4pt]
&&\phantom{\Gamma\quad}
= S_{\rm cl} + \frac 43\frac{q^2}{(4\pi)^2\epsilon} \, \cL_{1,A}
+  \frac{2q^2}{(4\pi)^2\epsilon} \, \cL_{1,\varphi}
\nonumber   \\[4pt]
&&\phantom{\Gamma \quad}
 +\frac{5\rm i}{12}\frac{q^3}{(4\pi)^2\epsilon}\,
 \theta^{\mu\nu}F_{\m\n}\bar\varphi{\bar\sigma}^\r(D_\r \varphi)
 - \frac{5}{6}\frac{q^3}{(4\pi)^2\epsilon}\, \theta^{\mu\nu}
F^{\rho\sigma}\ve_{\m\n\r\s}\bar\varphi{\bar\sigma}^\tau
(D_\tau\varphi)
 \hc   \, .
 \nonumber
\end{eqnarray}
The last two shifts, 
$\,{\bf A}_\rho^{\prime \prime} \,$ (which is a gauge transformation)
and $\,{\bf \Phi}^{\prime\prime} $,
\begin{eqnarray}
 && {\bf A}_\rho^{\prime \prime}=
-\frac{5}{6}\frac{q^2}{(4\pi)^2\epsilon}
 \, \theta^{\m\n}\ve_{\m\n\tau\s}(\p_\r F^{\tau\s}),\\[6pt]
&&
 {\bf \Phi}^{\prime\prime}= -\frac{5}{12}\frac{q^3}{(4\pi)^2\epsilon}\,\theta^{\mu\nu}F_{\m\n}\varphi
 -\frac{5\rm i}{6} \frac{q^3}{(4\pi)^2\epsilon}\, \theta^{\mu\nu} \ve_{\m\n\r\s}
F^{\rho\sigma}\varphi ,\nonumber
\end{eqnarray}
transform the effective action to
\begin{equation}
\Gamma = S_{\rm cl} + \frac 43\frac{q^2}{(4\pi)^2\epsilon} \, \cL_{1,A}
+  \frac{2q^2}{(4\pi)^2\epsilon} \, \cL_{1,\varphi}\, .                   \label{***}
\end{equation}
The remaining divergence  can  be removed by a multiplicative renormalization of
fields and coupling constants;  noncommutativity parameter $\theta$
 in principle gets renormalized, too.

\initiate
\section{Discussion}

We have seen in the previous section that it is possible to find a
Seiberg-Witten redefinition of commutative fields $A_\rho$ and $\varphi$
\begin{equation}
A_\rho\to A_\rho +{\bf A}_\rho +{\bf A}_\rho^{\prime}+{\bf A}_\rho^{\prime\prime}, \qquad
\varphi \to \varphi +{\bf \Phi} +{\bf \Phi}^\prime +{\bf \Phi}^{\prime\prime},
\end{equation}
which cancels all divergences in the one-loop correction to the effective action $\Gamma^{(1)}$,
except for classical interaction terms. This redefinition changes neither the physical
noncommutative theory nor its commutative limit; it changes only the identification of
the $\theta$-linear part of the action in terms of commutative fields $A_\r$ and $\varphi$.
This means that, had we taken instead of the simplest expansions
(\ref{expansion:A}-\ref{expansion:psi}), the other  defined by
\begin{eqnarray}
&&
\hat A_\rho = A_\rho  +\frac 14 q \, \theta^{\mu\nu} \{ A_\mu,
\partial_\nu A_\rho   + F_{\nu \rho} \} + a {\bf A}_\rho + a^\prime {\bf A}_\rho^{\prime}
+ a^{\prime\prime} {\bf A}_\rho^{\prime\prime} ,
\label{expA}
\\[4pt]
&&
\hat\varphi = \varphi
+\frac{1}{2}q\, \theta^{\mu\nu} A_\mu \partial_\nu\varphi
+b {\bf \Phi} + b^\prime {\bf \Phi}^\prime + b^{\prime\prime} {\bf \Phi}^{\prime\prime},
\label{exppsi}
\end{eqnarray}
we would have obtained  one-loop renormalizable action of the form\footnote{
The discussion here is confined only to divergences obtained in perturbation theory; the
question of chiral anomalies has to be treated additionally, see~\cite{Banerjee:2001un,Martin:2002nr}.}
\begin{eqnarray}
 &&\cL_{\rm NC} =\cL_{\rm C} +\kappa_1 \cL_{1,A} + \kappa_2 \cL_{1,\varphi}
\nonumber\\[4pt]
&&\phantom{\cL_{NC} =}
 +\kappa_3 \theta^{\m\n}\ve_{\m}{}^{\r\s\t} F_{\r\s} (D^2 F_{\n\t})
+{\rm i}\kappa_4 \theta^{\m\n} \Big({\rm i}\bar\varphi {\bar\s}_\r \s_{\m\n}(D^\r D^2\varphi) \hc \Big)
\nonumber\\[5pt]
&&\phantom{\cL_{NC} =}
+ \theta^{\m\n}\bar\varphi \Big( (\kappa_5 F_{\m\n} +\kappa_6 \ve_{\m\n\r\s}F^{\r\s}
+ \kappa_7 F_\m{}^\r\s_{\n\r} )\varphi \hc \,  \Big) .
\label{renL}
\end{eqnarray}
To prove the last statement rigorously one should in fact start
with (\ref{renL}), repeat all steps of quantization,
renormalize couplings $\, \kappa_i $ and $\theta^{\m\n}$ explicitly,
find $\beta$-functions, etc. This we will do in our following work. However, already
calculations  presented here strongly indicate renormalizability  because, due to
various identities, all divergent terms of appropriate dimension which could be 
obtained are already  included in (\ref{renL}).

In comparison with (\ref{lag}), lagrangian (\ref{renL}) contains new
interaction vertices: these are terms proportional to $\kappa_5$, $\kappa_6$
and $\kappa_7$. It contains also a modification of 
propagators. The change of the photon dispersion relation is perhaps more interesting
 because one hopes to compare its effects with the data on anisotropy
and polarization of the CMB radiation. In fact a comprehensive analysis of
various modifications of the photon dispersion relation was done already
in \cite{Kostelecky:2009zp}, and the term
$\, \kappa_3 \theta^{\mu\n}\varepsilon_{\mu}{}^{\rho\sigma\t} F_{\rho\sigma} \Box F_{\n\t}\, $
which we obtain here was included. Let us shortly discuss it.
From the free-photon part of the effective action
\begin{equation}
  \int - \frac 14 F_{\mu\nu} F^{\mu\nu}
 + \kappa_3 \theta^{\mu\n}\varepsilon_{\mu}{}^{\rho\sigma\t} F_{\rho\sigma} \Box F_{\n\t}
 \label{A}
\end{equation}
we obtain the equation of motion
\begin{equation}
 \p^\alpha F_{\alpha\beta} -\kappa_3 \theta^{\m\n}\left( 2\ve_{\m\alpha\beta\s} \eta_{\r\n}+
\ve_{\m\r\s\beta}\eta_{\alpha\n}-\ve_{\m\rho\s\alpha}\eta_{\b\n}\right)\p^\alpha\Box F^{\r\s}=0\, .
\label{E}
\end{equation}
Comparing (\ref{E}) to equations and to notation of \cite{Kostelecky:2007zz} we can identify
\begin{equation}
 (k_F)_{\beta\alpha\r\s} = \kappa_3 \theta^{\m\n}(\ve_{\m\alpha\b\s}\eta_{\n\r}
-\ve_{\m\alpha\b\r}\eta_{\n\s} +
\ve_{\m\rho\s\beta}\eta_{\n\a} -
\ve_{\m\rho\s\a}\eta_{\n\beta}) .
\end{equation}
It is easy to see however that due to identity
 (\ref{eps-del}),  $k_F $ vanishes! Therefore in fact the additional $\theta$-linear
term does not change the propagation of free photons: they satisfy  the Maxwell equations,
$\,  \p^\alpha F_{\alpha\beta}= 0\, $.
There is no vacuum birefringence of photons, that is, none in linear order in
$\theta$. Were it present, the comparison with the observational data done 
in~\cite{Kostelecky:2007zz}
would give that the scale of noncommutativity is of order 30 TeV\footnote{That is,
$\, k_F \sim \Lambda_{{\rm NC}}^{-2} \sim 10^{-9} {\rm GeV}^{-2}$.},
which is  roughly in agreement with
previously obtained constraints \cite{Minkowski:2003jg,Horvat:2009cm,Horvat:2010sr}.

Spinors however behave differently.
The modified free spinor action
\begin{equation}
\int {\rm i}\bar\varphi \left({\bar\s}_\r  \p^\r
+ {\rm i}\kappa_4 \theta^{\m\n} {\bar\s}_\r \s_{\m\n}\p^\r \Box \right) \varphi \hc\,,
\end{equation}
implies the equation
\begin{equation}
({\rm i}\bar\sigma^\r\partial_\r+{\rm i}\kappa_4\theta^{\m\n}\ve_{\m\n\r\s}\bar\s^\s\partial^\r \Box)\varphi =0 ,
\label{EE}
\end{equation}
and we easily see that in this case the dispersion changes.
Let us assume that noncommutativity
is spatial, $\theta^{0i} =0$,
and denote $\,(\theta^{12})^2 = \theta^2_\perp$,\
$\,(\theta^{13})^2+(\theta^{23})^2 = \theta^2_\parallel \,$;
 the momentum is along the third axis,   $ \,k^\mu = (E, 0, 0, p)\,$.
The dispersion  relation  becomes
\begin{equation}
 k^2 \Big(1 -
4\kappa_4^2\theta^2_\parallel \, p^2 k^2-4\kappa_4^2(\theta^2_\perp+\theta^2_\parallel)\, k^4\Big)=0
\end{equation}
and has the solutions
\begin{equation}
k^2 =0 \,, \qquad
 k^2  = \frac{ \sqrt{\frac{1}{\kappa_4^2}(\theta^2_\perp + \theta^2_\parallel)
+ \theta^4_\parallel \, p^4}- \theta^2_\parallel \, p^2 }
{2 (\theta^2_\perp + \theta^2_\parallel)} \ .
\end{equation}
One of the  propagating fermionic modes acquires mass 
which is  for small noncommutativity
very large, of order $1/\sqrt{\theta}$, and thus on  cosmological distances
 it is effectively  supressed.  As the mass 
depends on the direction of propagation with respect to  noncommutativity 
$\theta^{\m\n}$ this mode is birefringent.

The possibility of photon birefringence due to noncommutativity was
first discussed in \cite{Carroll:2001ws} within a classical
$\theta$-expanded gauge model. It was obtained that the effect
exists in linear order only if there is an external electromagnetic
field, otherwise the birefringence is of second order in $\theta$. This
result was expanded in~\cite{Mariz:2006kp}. Here also the first-order 
birefringence of photons exists in the external field but not in vacuum. Modifications 
of the photon propagator due to quantum corrections were thoroughly 
analyzed in many papers within the non-expanded noncommutative
$\mathrm{U}(1)$
theory,~\cite{Brandt:2002if,Zahn:2006mg,Abel:2006wj}. However, as
the theory is not perturbatively renormalizable it is  not clear
how to interpret the quantum corrections and to relate them to
observations \cite{Horvat:2010km}. The analysis within a
nonperturbative numerical approach was done in
\cite{Bietenholz:2008tp}.

On the other hand, birefringence of chiral fermions obtained here
is a completely new effect:
it is absent for Dirac particles,~\cite{Wulkenhaar:2001sq,Buric:2002gm}.
As astrophysical effects related to fermion propagation, for example for neutrinos,
are very weak, it is  not clear whether such effect can be tested experimentally in
astrophysical measurements; perhaps high energy experiments would prove
better for this task.
In any case, physical implications of the obtained model need to be analyzed in more
details and we plan to study them in our future work.

Therefore perhaps the main importance of the presented result is that 
another noncommutative gauge model with good renormalizability 
properties is found, 
and that it can be used as a building block for constructing 
noncommutative generalizations of the Standard Model.
As we mentioned, a class of such models was found in
 \cite{Martin:2009sg,Martin:2009vg,Tamarit:2009iy}.
In these papers, requirement of renormalizability (at one loop, in 
$\theta$-linear order 
and on-shell)  singled out  GUT-compatible and anomaly-safe 
$\theta$-expanded theories. Technically, these are the 
theories in which the left-handed and the conjugate of the right-handed 
fermion are in the same  representation of the gauge group.
Renormalizability implied that the triple gauge boson 
interactions were absent.
Our model is somewhat different: it includes only one, say left-handed
fermion;  renormalizability is also one-loop and $\theta$-linear but off-shell,
and the triple gauge-boson interactions are a priori allowed. Our framework
is less resticting; however to achieve renormalizability we need the 
Seiberg-Witten redefinition of all fields. 

But also in our model one can see that the GUT-compatibility is an 
important requirement. Let us 
 assume that besides the left-handed spinor $\varphi$
we also have  a right-handed spinor ${\bar \chi}^\pm$
of the same or of the opposite charge\footnote{Notation
$\chi^\pm$ is taken to be in accordance with~\cite{Martin:2009sg}. }.  
Repeating the calculations for the one-loop correction of the 
gauge field propagator we  obtain
\begin{equation}
{\Gamma}_{2,A} =\frac{1\pm 1}{12}\frac{q^2}{(4\pi)^2\epsilon}\theta^{\mu\nu} 
{\varepsilon_{\mu\rho\sigma\tau}}
(\partial_\lambda F^{\rho\lambda})(\partial_\nu F^{\sigma\tau} ) ,         \label{23}
\end{equation}
while the divergence of the fermion propagators is given by
\begin{equation}
{\Gamma}_{2,\varphi} =
- \frac{1}{12} \frac{q^2}{(4\pi)^2\epsilon}
 \theta^{\mu\nu}
\varepsilon_{\mu\nu\rho\sigma}\Big( \, 
{\rm i} (D^\rho\bar\varphi){\bar\sigma}^\sigma
(D^2\varphi)  \pm {\rm i} 
 (D^\rho\bar\chi^\pm){\bar\sigma}^\sigma
(D^2\chi^\pm) 
\hc \Big)   .                                    \label{24}
\end{equation}
In both results of course both bosons and fermions run in the loop.
The difference in signs in (\ref{23}-\ref{24})
comes from the fact that the action
for the GUT-compatible spinor $\chi^-$ differs from
 the action for the $\chi^+$ by the change $\theta\to-\theta$, \cite{Martin:2009sg}.
Therefore if the model contains the pair ($\varphi$,$\chi^-$),
the bosonic divergence vanishes. Analogously, it is easy to see
that all gauge field redefinitions vanish too.
This emphasises the fact that divergent term ${\Gamma}_{2,A}$
comes from, and depends on the fermion-boson interaction, and in 
specific cases the fermion loops cancel. For this reason
also in the case of pure gauge U(1) and SU(N) theory
there is no gauge field redefinition at linear 
order,~\cite{Bichl:2001cq,Buric:2005xe}. 

The present result  opens new perspectives, while some of the old questions remain.
The first and perhaps really nontrivial one is whether the field redefinitions
are enough to ensure renormalizability also in  quadratic order in $\theta$.
Though this question is technically very hard, it could happen that some
additional Ward identities can help to resolve it, and we hope that this issue will be 
addressed and clarified in the future.

\vspace{.5cm}

{\bf Acknowledgement}
We would like to thank M. Dimitrijevi\' c for pointing out 
Schouten  $\ve$-$\eta$ identity (\ref{eps-del}).
The work of M.~B., V.~R. and D.~L. is a done within the project
171031 of the Serbian Ministry of Science. The work of J.~T. is
supported by the project 098-0982930-2900 of the Croatian Ministry
of Science, Education and Sports and in part by the EU (HEPTOOLS) project
under contract MRTN-CT-2006-035505.

\initiate
\appendix
\section{Conventions}

 We use the following chiral representation
of $\gamma$-matrices
\begin{equation}
\gamma^\mu =
\begin{pmatrix}
0 & \sigma^\mu \\
\bar\sigma^\mu & 0
\end{pmatrix},
\quad
\gamma_5 =
\begin{pmatrix}
-1 & 0\\
0 & 1
\end{pmatrix},\;\;
\end{equation}
with
\begin{equation}
\sigma^\mu = (1,\vec\sigma), \quad \bar\sigma^\mu = (1,
-\vec\sigma)                   \label{gama}
\end{equation}
and
\begin{equation}
 \s_{\m\n} = \frac 14 (\s_\m\bar\s_\n - \s_\n \bar\s_\m),\quad
 \bar\s_{\m\n} = \frac 14 (\bar\s_\m\s_\n - \bar\s_\n \s_\m) .
\end{equation}
In the field redefinitions we use
\begin{eqnarray}
&& -\mathrm{i}\ve_{\m\n\r\t}{\bar\sigma}^\t = {\bar \sigma}_\m \sigma_\n {\bar\sigma}_\r
+\eta_{\m\r}{\bar\sigma}_\n -\eta_{\n\r}{\bar\sigma}_\m  - \eta_{\m\n}{\bar\sigma}_\r,
\label{3sigma}
\\[8pt]
&& \bar\s^\m \sigma^{\n\r} +\bar\s^{\n\r}\bar\s^\m = - \mathrm{i} \ve^{\m\n\r\t}\bar\s_\t
\end{eqnarray}

Chiral spinors  $\varphi$, $\chi$ multiply as
\begin{equation}
\varphi \chi =\chi\varphi,\quad \bar\varphi \bar\chi
=\bar\chi\bar\varphi, \label{ind}
\end{equation}
\begin{equation}
\bar\varphi\bar\sigma^\mu \chi =-\chi\sigma^\mu\bar\varphi~, \quad
(\chi\sigma^\mu\bar\varphi )^\dagger =\varphi\sigma^\mu\bar\chi~.
\label{inc}
\end{equation}
Those relations, as can be seen easily, give the usual identities
for  Majorana spinors $\phi$, $\psi$
\begin{equation}
\bar\phi \psi =\bar\psi \phi~,\quad \bar\phi\gamma_5 \psi =\bar\psi
\gamma_5\phi~,\nonumber
\end{equation}
\begin{equation}
\bar\phi \gamma^\mu \psi = -\bar\psi \gamma^\mu \phi~,\quad \bar\phi
\gamma^\mu\gamma_5\psi =\bar\psi\gamma^\mu\gamma_5 \phi~.
\end{equation}
Majorana lagrangians are obtained from the corresponding chiral
ones using identities~(\ref{ind}-\ref{inc}) and hermiticity of the lagrangian.

Finally relation between $\ve$ and $\eta$ tensors (Schouten identity) reads
\begin{equation}
 \ve_{\m\n\r\s}\eta_{\t\l} +\ve_{\m\n\t\r}\eta_{\s\l} -\ve_{\m\n\t\s}\eta_{\r\l}
 +\ve_{\t\m\r\s}\eta_{\n\l}-\ve_{\t\n\r\s}\eta_{\m\l} = 0\, .
\label{eps-del}
\end{equation}
 Multiplying (\ref{eps-del}) by $\theta^{\m\n} F^{\r\s}D^\lambda$
we obtain another useful relation
\begin{equation}
 \theta^{\m\n} F^{\r\s} \left(2 \ve_{\m\r\s\t}D_{\n} + 2\ve_{\m\n\r\t}D_{\s}
 -\ve_{\m\n\r\s}D_{\t} \right)  = 0\, .
\label{e-d}
\end{equation}

\end{document}